# On measuring the size of nuclei of comets

*S. N. Dolya*

*Joint Institute for Nuclear Research, ul. Joliot-Curie 6, Dubna, 141980, Russia*

**Abstract**

Possibilities of measuring the size of nuclei of comets hidden by dust clouds are discussed. To this end, the dust cloud should be irradiated with a flow of rods accelerated in a linear mass accelerator to the velocity $V_{fin}$ = 6 km / s. Each rod should be equipped with a transmitter with a power of 1 µW, which is destroyed in a collision with a comet's nucleus, or continues to work if the rod passes through the dust cloud without collision. Radio signals are received by three independent ground stations. At a distance of R = 1000 km from the nucleus of the comet the power of the received signals is $P_{sig}$ = $10^{-17}$ W, the receiver noise power: $P_{noise}$ = $10^{-20}$ W.

**Introduction**

To measure the size of nuclei of comets is very important since a comet, even flying at a considerable distance from the Earth, can cause serious damage.

Let us consider in more detail the possibility of exposing the dust cloud that hides the nucleus of the comet to the flow of rods, each equipped with a radio transmitter.

**1. Acceleration of rods**

*1. 1. Parameters of the accelerated rod*

We will consider acceleration of macroparticles shaped as a rod with a conical head, which are electrically charged.

Acceleration of rods in a helical waveguide is well studied [1]. This acceleration requires that the initial velocity of the rod and the phase velocity of the wave were approximately identical. As the acceleration of rods goes on, the wave phase velocity in the spiral waveguide should be increased so that rod was always in the same phase of the wave, called a synchronous phase. The phase velocity of the wave in the waveguide can be increased by increasing the coil pitch or decreasing the coil radius, or doing both at the same time.

Let the diameter of the rod be $d_{sh} = 2$ mm and length $l_{sh} = 300$ mm. Then the cross-section area of the rod is $S_{tr} = \pi d_{sh}^2 / 4 = 3.14 * 10^{-2}$ cm$^2$, and the volume of the rod is $V_{sh} \approx 1$ cm$^3$. The mass of the rod at an average density $\rho_{aver} = 5$ g/cm$^3$ is $m_{sh} = 5$ g.

*1. 2. The ratio Z / A*

We assume that the average atomic mass of the rod is $A_{sh} = 30$. The number of nucleons in the rod can be found from the proportion

$$6 * 10^{23} \text{ -------- } 30 \text{ g}$$
$$x \text{ ---------- } 5 \text{ g}$$

where $x = 10^{23}$ atoms or $A_{sh} = 3 * 10^{24}$ nucleons.

We take the surface tension of the electric field on the rod to be $E_{surf} = 3 * 10^8$ V / cm. Using the formula for the surface tension of the field for the cylinder

$$E_{surf} = 2\kappa / r, \qquad (1)$$

we find the charge density per unit length of the rod

$$\kappa = E_{surf} * r/2e = (5 * 10^7 * 0.1) / (5 * 10^{-10} * 300 * 2) = 10^{14}, \qquad (2)$$

from which we can obtain

$$N_p = (\kappa / e) * l_{sh} = 3 * 10^{15}. \qquad (3)$$

Thus, if $N_p = 3*10^{15}$ protons are set on the rod, the surface tension of the field will turn out to be $E_{surf} = 3 *10^8$ V / cm.

Now that we know the total number of excess protons on the rod $N_p = 3 * 10^{15}$ and the number of nucleons in it $A_{sh} = 3 * 10^{24}$ we can find the charge-to-mass ratio for the rod $Z / A = N_p / A = 3 * 10^{15}/3 * 10^{24} = 10^{-9}$.

*1.3. Proton beam irradiation of rods*

To accelerate a rod shaped as a cylindrical rod with a pointed-cone head in a spiral waveguide, it should be electrically charged. The electric charge can be imparted to a rod by irradiating it with a beam of protons so that the irradiating protons remained on it. Then the electric charge of rods will increase in proportion to the proton beam current and the duration of exposure. Let the proton beam current be $I_{beam} = 0.5$ A, and the current pulse duration be $\tau_{beam} = 1$ ms. Then the total number of protons in the current pulse is $N_p = I_{beam} * \tau_{beam} / e = 3 * 10^{15}$ protons.

*1.4. Proton beam irradiation of rods. Proton energy*

Let a cylindrical rod gas-dynamically accelerated to $V_{in} = 1$ km / s is exposed to a proton beam from an external source. The field surface strength is assumed to be $E_{surf} = 300$ MV / cm. Then, for the diameter of the cylinder $d_{sh} = 2$ mm, we find that the minimum energy of the protons that can overcome the Coulomb repulsion of the protons previously placed on the rod should be $W_p > eE_{surf} * d_{sh} / 2 = 30$ MeV.

*1. 5 . Proton beam irradiation. Mean free path*

Protons with an energy of 30 MeV have a range in aluminum about 1 g/cm$^2$ [2, p. 953]. Given the density of aluminum $\rho_{Al} = 2.7$ g/cm$^3$, we find that the range of the protons in aluminum is $l_{Al} \approx 3$ mm. Since the average density of the material we have chosen for the cylinder is $\rho_{aver} = 5$ g/cm$^3$, about twice the density of aluminum, the mean free path of protons with an energy of 30 MeV in the rod will be approximately 2 mm.

Apparently, it is necessary to gradually increase the energy of the protons during the irradiation. As more and more protons are set on the rod, the energy of the protons emitted later, on the one hand, should be sufficiently high to overcome the Coulomb repulsion of the protons that are already on the rod and, on the other hand, should be such that the path of the protons in the material of the rod was much smaller than its diameter.

In this energy range the path of the protons in the material increases linearly with energy; for example, protons with energy $W_p = 3$ MeV, has a path $3 * 10^{-2}$ g/cm$^2$ [2, p. 953], or about 100 µ, and will not be able to cross the rod diameter of 2 mm. They will lose their energy by ionization of the material and would stay within the rods.

## 2. Acceleration length

The acceleration rate of a charge in an electric field can be written as

$$\Delta W = (Z / A)\, eE_{zw}, \qquad (4)$$

and for the strength of the wave $E_{zw} = 70$ kV / cm, the rate of the energy gain will be $\Delta W = 7 * 10^{-4}$ eV / (m * nucleon), so that the required increase in energy $\Delta \varepsilon = 0.2$ eV / nucleon will be attained over the length

$$L_{acc} \approx \Delta \varepsilon / \Delta W = 30 \text{ m}. \qquad (5)$$

## 3. Selection of the spiral waveguide parameters

The spiral waveguide is a standard coaxial cable with its central wire wound into a spiral. In such a cable, there is no dispersion in a wide range of frequencies, i.e., the velocity of propagation does not depend on the frequency and the phase velocity in this cable coincides with the group velocity.

The wave (pulse) propagation velocity V in this cable is determined by the tightness of winding of the central conductor into a spiral and the dielectric properties of the medium that fills the cable. This relation is called the dispersion equation

$$\beta = \text{tg}\, \Psi / \varepsilon^{1/2}, \qquad (6)$$

where $\beta = V / c$, V is the velocity of the pulse through the cable, $c = 3 * 10^{10}$ cm / s is the speed of propagation of electromagnetic waves in a vacuum, tg $\Psi = h/2\pi r_0$, h is the winding pitch of the spiral, $r_0$ is the radius of the spiral winding, $\varepsilon$ it the relative dielectric constant of the medium filling the cable. The wave as is runs along the spiral in a circle of $2\pi r_0$, while moving a small distance h along the axis of the spiral. The wave further slows down due to the dielectric properties of the medium determined by the value of $\varepsilon$.

To accelerate a body by a pulse running in a cable, the insulator must be removed from inside the spiral, and then the pulse velocity in this cable will slightly increase [1] ,

$$\beta = \sqrt{2} * \text{tg } \Psi/\varepsilon^{1/2}. \qquad (7)$$

The pulse running through a line with distributed parameters contains not only the gradient of the magnetic field, which are accelerated magnetic dipoles, but also the electric field $E_{zw}$, which can accelerate a charged body.

The initial velocity of the projectile in a spiral $\beta_{sh\,in}$ expressed in terms of the speed of light $\beta_{in} = V_{in}/c$, where $c = 3*10^{10}$ cm / s is the speed of light in a vacuum, is $\beta_{in} = 3.3 * 10^{-6}$, and the final velocity is $\beta_{fin} = 2 * 10^{-5}$. Spiral should apparently consist of several sections, so that an optimal acceleration rate could be selected within each section. The acceleration wavelength can be determined from the condition $x = 2\pi r_0 / (\beta * \lambda_0) = 1$, where x is a dimensionless parameter in the arguments of the modified Bessel functions, $r_0$ is the radius of the spiral, $\beta$ is the phase velocity, $\lambda_0$ is the acceleration wavelength in a vacuum, and $\lambda_0 = s/f_0$, $f_0$ is the acceleration frequency.

With the initial radius of the spiral equal $r_{0\,in} = 20$ cm, and the dielectric constant of the medium between the spiral and the screen $\varepsilon = 1280$, we find: $\lambda_0 = 3.8 * 10^7$ cm, $f_0 = 790$ Hz. Thus, the slow wavelength for the beginning of the acceleration is $\lambda_{slow} = \beta\lambda_0 = 1.25$ m.

*3.1. Parameters of the spiral*

In order to obtain the required field intensity $E_0$ in the spiral waveguide, we need a power to be introduced into it, which is defined by the formula [1]

$$P = (c/8) * E_0^2 * r_0^2 * \beta * \{\}, \qquad (8)$$

where P is the high-frequency power introduced into the coiled waveguide, $r_0$ is the radius of the spiral, and $\beta$ is the phase velocity of the wave, which is determined from the dispersion equation. The braces in (8) are

$$\{\} = \{(1 + I_0 K_1/I_1 K_0)(I_1^2 - I_0 I_2) + \varepsilon (I_0/K_0)^2 (1 + I_1 K_0/I_0 K_1)(K_0 K_2 - K_1^2)\}, \qquad (9)$$

where $I_0$, $I_1$, $I_2$ are the modified Bessel functions of the first kind, $K_0$, $K_1$, $K_2$ are the modified Bessel functions of the second kind. The first term in the braces corresponds to the flow propagating inside the spiral, and the second term corresponds to the flux traveling outside the spiral. Since the space between the spiral and the screen is filled with a dielectric, a factor $\varepsilon$ appears in front of the second term [1].

In this case, deceleration of the electromagnetic wave to velocities of the order of the velocity of sound requires the use of both the geometrical properties of the structure (small-pitch spiral) and the properties of the medium, for which we chose the relative permittivity $\varepsilon = 1280$.

Thus, the flow of the high-frequency power propagating outside the spiral is more than $10^3$ times higher than the power that propagates inside the spiral. Therefore, the first term inside the braces can be neglected, and the value of the braces for the argument $x = 1$ is approximately $\{\} \approx 4\varepsilon$.

In accelerators the synchronous phase is selected on the front slope of the pulse, so that the electric field accelerating the rod is always lower than the peak value. Let us choose the synchronous phase $\varphi_s = 45^0$, $\sin \varphi_s = 0.7$, $E_{zw} = E_0 \sin\varphi_s$. Thus, the amplitude of the wave which accelerates the cylindrical rod should be

$$E_0 = E_{zw} / \sin \varphi_s = 100 \text{ kV} / \text{cm}. \qquad (10)$$

Then, the wave power in watts expressed by the formula (8) is
$$P \text{ (W)} = 3 * 10^{10} * 10^{10} * 10^2 * 4 * 3.3 * 10^{-6} * 1.28 * 10^3 * 4 / /(8 * 9 * 10^4 * 10^7) = 300 \text{ MW}. \qquad (11)$$

*3.2. Transition from a sine wave to a single pulse*

This power is achievable for pulse technology. We expand the sinusoidal pulse [1], corresponding to the half-wave $E_{pulse} = E_{0pulse}\sin(2\pi t/T_0)$, $2\pi/T_0 = \omega_0$, $\omega_0 = 2\pi f_0$ in a Fourier series.

$$f_1(\omega) = (2/\pi)^{1/2} \int_0^{T_0/2} \sin\omega_0 t * \sin\omega t \, dt. \qquad (12)$$

The pulse spectrum is narrow and covers the frequency range from 0 to $2\omega_0$. Since the spiral waveguide dispersion (dependence of the phase velocity on the frequency) is weak, it can be expected that the full range of frequencies from 0 to $2\omega_0$ will propagate at an approximately identical phase velocity. As a result, the half-wave sinusoidal pulse will spread out several-fold in space due to only an increase in the phase velocity of the wave. In this case the spiral waveguide should be matched with the supply feeder in the band $\Delta f \approx \omega_0/2\pi$.

We introduce the concept of the pulse amplitude $\tilde{U}$ connected with the field strength in the axis of the spiral $E_0$ by the relation [1],

$$\tilde{U}_{pulse} = E_{0pulse}\lambda_{slow}/2\pi, \quad \lambda_{slow} = \beta\lambda_0, \quad \lambda_0 = c/f_0. \qquad (13)$$

The wavelength choice $\lambda_0 = 3.8 * 10^7$ cm means, that we choose the duration of the acceleration of the rod ($f_0 = c/\lambda_0 = 790$ Hz), $\tau_{pulse} = 1 / (2f_0) = 630$ μs. The voltage pulse amplitude will be $\tilde{U} = E_0\lambda_{slow} / 2\pi = 2$ MV, and the pulse current through the coil windings will be $\tilde{I} = P / \tilde{U} = 150$ A. Table 1 summarizes the main parameters of the accelerator.

Table 1. Parameters of the accelerator

| | |
|---|---|
| $Z/A = 10^{-9}$, dielectric outside spiral, wave power, P | P = 300 MW |
| | μ = 1, ε = 1280 |
| Velocity, initial – final, $\beta_{ph}$ | $\beta_{ph} = 3.3*10^{-6} - 2*10^{-5}$ |
| Initial radius of spiral, $r_0$ | $r_0 = 20$ cm |
| Wave frequency, $f_0$, | $f_0 = 790$ Hz |
| Electric field strength, $E_0$ | $E_0 = 100$ kV/cm |
| Accelerator length, $L_{acc}$ | $L_{acc} = 30$ m |
| Pulse duration, τ | τ = 630 μs |
| Voltage amplitude, $\tilde{U}_a$ | $\tilde{U}_a = 2$ MV |

*3. 3. The capture of rods in the acceleration mode. Admission*

Let us calculate the required accuracy to match the initial phase of the accelerating wave (pulse) with the synchronous phase. The theory of rod capture into a traveling wave gives [3], $\Delta\varphi = 3\varphi_s$, (+ $\varphi_s$ - $2\varphi_s$). In practice, it means, for example, that in our case, where T / 4 corresponds to the duration of 316 μs or $90^0$ degrees, one phase corresponds to the time interval of approximately 3 μs.

In linear accelerators the buncher gives bunches with the phase width $\pm 15^0$. To avoid large phase fluctuations, we require that the accuracy of synchronization of rods with the accelerating pulse be $\Delta\tau = \pm 15 * 3$ μs= $= \pm 45$ μs. This synchronization accuracy seems quite attainable for the gun powder start, that is, preliminary gas-dynamic acceleration of rods.

Let us now calculate the accuracy tolerance for coincidence of the initial rod velocity and the phase velocity of the pulse propagating along the helical structure. We introduce the quantity $g = (p-p_s) / p_s$, which is the relative difference of the pulses [3]. In the non-relativistic case is simply a relative velocity dispersion of $g = (V-V_s) / V_s$. The vertical scale of the separatrix is calculated by the formula [3]

$$g_{max} = \pm 2 [(W_\lambda ctg\varphi_s / 2\pi\beta_s) * (1 - \varphi_s / ctg\varphi_s)]^{1/2}, \qquad (14)$$

where $\varphi_s = 45^0 = \pi / 4$, $ctg\varphi_s = 1$ $[1 - \varphi_s / ctg\varphi_s]^{1/2} = 0.46$ $2 * 0.46 = 0.9$, and $W_\lambda = (Z / A) eE_0\lambda_0 sin\varphi_s / Mc^2$.

Let us find $W_\lambda = (Z / A) eE_0\lambda_0 sin\varphi_s / Mc^2$, the relative energy gain of the rod over the wavelength $\lambda_0$ in vacuum. In our case $\lambda_0 = c/f_0 = = 3.8 * 10^7$ cm, $sin\varphi_s = 0.7$, $Mc^2 = 1$ GeV, and $W_\lambda = 2.66 * 10^{-5}$. Substituting numerical values, we get $g = (V_{in}-V_s) / V_s = \Delta V / V_s$, and, finally, $\Delta V / V_s = \pm [2.66 * 10^{-5} / (6.28 * 3.3 * 10^{-6})]^{1/2} * 0.9 = \pm 0.3$.

Thus, the tolerable discrepancy between the initial rod velocity and the pulse velocity is of the order of $\Delta V / V_s = \pm 30\%$. For the initial rod velocity $V_{in} = 1$ km / s of the tolerable velocity deviation is $\Delta V < 300$ m / s.

**4. Radial motion**

As is well known [3], in azimuth - symmetric wave of phase stability, the phase stability region corresponds to the radial defocusing. In this case, we cannot have simultaneously radial and phase stability; under the phase stability conditions external field are required for radial focusing. In this phase regions the radial component of the electric field of the wave is directed towards the increasing radius, i.e., radially accelerates the rods.

In this region of rod velocities, hypersonic, hundreds of thousands of times lower than the speed of light, focusing by magnetic quadrupole lenses is not effective, and electrostatic quadrupole lenses are most suitable for this purpose. They focus the rods in one plane and defocus them in another. Collected into a doublet, two such lenses give the resulting focusing effect. The accelerator should be divided into separate sections, and the focusing doublets can be placed between the acceleration sections.

## 5. Carrying rods away beyond the Earth's atmosphere

### 5.1. Lift

When the length of the accelerator is $L_{acc} \approx 30$ m, it should be placed horizontally. To carry a rod beyond the atmosphere, we can use a small asymmetry in the rod's shape, such that it creates a lifting force $F_y$. The relevant equation of vertical motion can be written as

$$mdV_y / dt = C_y \rho_0 V_x^2 * S_{tr} / 2, \qquad (15)$$

where $C_y$ is the aerodynamic lift coefficient, $\rho_0 = 1.3 * 10^{-3}$ g/cm$^3$ is the air density at the surface of the Earth, $V_x$ is the horizontal velocity of the rod, and $S_{tr}$ is the cross-section of the rod.

### 5.2. Ballistics. Air resistance

We calculate the motion of an electrodynamically accelerated rod. Equations of motion of the rod can be written as

$$mdV / dt = - \rho C_x S_{tr} V^2 / 2, \qquad (16)$$

where m is the mass of the rods, V is the velocity, $\rho = \rho_0 e^{-z}/H_0$ is the barometric formula for changes in atmospheric density with height, $\rho_0 = 1.3 * 10^{-3}$ g/cm$^3$ is the air density at the surface of the Earth, $H_0 = 7$ km is the height in which the density decreases by e times.

The aerodynamic coefficient or the coefficient of drag is a dimensionless quantity that takes into account the "quality" form of rods,

$$C_x = F_x / (½) \rho_0 V_0^2 S_{tr}. \qquad (17)$$

Equation (16) can be written as

$$V(t) = V_0 / [1 + \rho C_x V_0 * S_{tr} * t/2m]. \qquad (18)$$

To calculate the rate of change of magnetic dipoles with time, it is necessary to find the aerodynamic coefficient $C_x$.

*5. 3. The calculation of the drag coefficient for air rods*

We assume that the rod is shaped as a cylindrical rod with a conical head. The impact of a nitrogen molecule at the sharp cone causes a change in the longitudinal velocity of the molecules

$$\Delta V_x = V_x * \Theta_h^2 / 2, \qquad (19)$$

where $\Theta_h$ is the angle at the vertex of the cone. The molecules of the nitrogen to pass rod pulse:

$$p = mV = \rho V_x S_{tr} t * \Delta V_x. \qquad (20)$$

The change in momentum per unit time - the power, the power of a frontal inhibition,

$$F_{x1} = (½) \rho V_x S_{tr} * V_x * \Theta_h^2. \qquad (21)$$

Dividing $F_{x1}$ by $(½) \rho V^2_x S_{tr}$, we get the drag coefficient for a sharp cone in the mirror image molecules from the cone (the Newton)

$$C_{x\,air} = \Theta_h^2. \qquad (22)$$

Let the length of the conical part of the rod be $l_{cone} = 12.5$ mm and the diameter be $d_{sh} = 2$ mm. This means that the angle at the vertex of the cone is $\Theta_t = d_{sh} / l_{cone} = 1.6 * 10^{-1}$ and $C_{x\,air} = 2.5 * 10^{-2}$.

In order to have a pointed-cone head, the rod should be sufficiently long. Limiting the length of the rods is the fact that for a good efficiency of the acceleration length $l_{tot}$ rods should be less than a quarter wavelength delayed $\lambda_{slow} = \beta\lambda_0$, i.e.: $l_{tot} < \beta\lambda_0 / 4$. In this case, for the begin of the acceleration, $\beta\lambda_0 / 4 = 40$ cm.

*5. 4. Rod passage through the atmosphere*

Let us draw up a table to show the time dependence of the vertical velocity, lift, and horizontal velocity of the rod. The vertical velocity is calculated by the formula

$$\Delta V_y = C_y \rho V_x^2 * S_{tr} * \Delta t / 2m. \qquad (23)$$

The climb is calculated by the formula

$$H_{fly\ n+1} = H_{fly\ n} + \tilde{V}_y * \Delta t + C_y \rho V_x^2 * S_{tr} * (\Delta t)^2 / 4m, \quad (24)$$

where $\tilde{V}_y$ is the average vertical velocity in the time interval $\Delta t$. A decrease in the horizontal velocity over time will be described by formula

$$V_{x\ n+1} = V_{x\ n} / [\ 1 + (C_x \rho V_{xn} * S_{tr} * \Delta t / 2m)]. \quad (25)$$

The change in the air density with altitude will be taken into account by the barometric formula $\rho = \rho_0 * \exp[-y/H_0]$, where $H_0 = 7$ km.

Table 2 shows the cylinder flight parameters as a function of time. The second column shows the vertical velocity of the cylinder, and the third shows the horizontal velocity of the cylinder, the fourth shows the height it gained after the corresponding second of flight, and the fifth shows the density of the atmosphere at this height.

Table 2. Flight parameters at $C_x, C_y = 2.5 * 10^{-2}$.

| t, s | $V_x$, km/s | $V_y$, km/s | Y, km | $\rho_{air}$, g/cm$^3$ |
|---|---|---|---|---|
| 0 | 6 | 0 | 0 | $1.3*10^{-3}$ |
| 10 | 3.72 | 3.67 | 18 | $2*10^{-4}$ |

In this case the time of gaining the maximum height is $\tau_{max} = V_y / g = 367$ s, where $g = 10^{-2}$ km/s$^2$ is the gravity acceleration, the flight range is $S = V_x * 2\tau_{max} = 2700$ km, and the maximum height is $Y = V_y^2/2g = 670$ km.

**6. Rod flight path control**

To control the flight path of the rod at its side surface is applied four quadrants of a material with different resonant absorption of laser radiation. "Right - left" and "up - down" deviations of the rod are effected by evaporating a corresponding quandrant exposed to laser radiation with a resonance wavelength.

In silicon with different degrees of doping the Langmuir frequencies $\omega_{pl}$ will be different and, therefore, these four quadrants will have different resonant frequencies of absorption [4].

Thus, we can assure that only one of the four quadrants, namely, that whose position is opposite to the direction from which the rod must be diverted, will evaporate upon resonant absorption of laser radiation and produce a jet.

In order to evaporate the quadrant in a short time, such that the heat from absorption of the laser radiation did not penetrate the body, the laser pulse must be sufficiently short.

*6. 1. Parameters of the flight and trajectory change*

Let the comet move as far from the surface of the Earth as $h_{com} = 200 - 400$ km and the distance from the comet to the floating platform carrying the accelerator and the laser is $s_2 = 1000$ km. Rod has a mass $m_b = 5$ g and moves with a velocity $V_b = 3$ km / s. When a mass $m_{jet} = 15$ mg moves at a velocity $V_{jet} = 1$ km / s perpendicular to the rod velocity, the transverse momentum transfer will be $p^\perp = m_{jet} * V_{jet}$, and this will result in the deflection angle $\theta^\perp = p^\perp / (m_b * V_b) = 10^{-3}$. This angle, on a distance from the target: $s_1 = 100$ km, will result in rejection of the body the trajectory: $\Delta l = s_1 * \theta^\perp = 100$ m.

*6. 2 . Energy relations for the jet efflux*

The silicon heat capacity is $c_{SI} = 20$ J / (mol * degree) [5, p. 199], melting point $T_{mel} = 1415\ ^0$C, solid - liquid phase transition heat $\Delta H_m = 50$ kJ / mol, boiling point $T_{eva} = 3300\ ^0$C, and liquid - vapor phase transition heat $\Delta H_m = 356$ kJ / mol [5 , p. 289]. Considering all energy needed for evaporation and the fact that 1 mole of silicon is 28 g, we find that evaporation of 1 gram of silicon requires an energy of ~ 15 kJ / g.

For the average directed velocity of silicon atoms to be $V_{jet} = 1$ km / s, the thermal velocity should be $V_T = 2.5$ km / s. Indeed, after averaging the velocity in one of the transverse plane, we get $\tilde{V}_1 = (V_T / \pi) * \int \sin\varphi\, d\varphi = = (2 / \pi) * V_T$, where integration over angles is from 0 to $\pi$. After averaging in two transverse planes we obtain $\tilde{V}_2 = V_{jet} = (2 / \pi)^2 * V_T \approx 0.4\ V_T$, so that , in addition to evaporation of silicon, it is necessary to impart thermal velocity $V_T = 2.5$ km / s to its atoms for their average directed velocity to be $V_{jet} = 1$ km / s.

We find the energy of the silicon atom moving with a velocity $V_T$ from the relation $m_{Si} * V_T^2 / 2 = \varepsilon_{Si} = 1.5 * 10^{-19}$ J. Given that 1 g contains $2 * 10^{22}$ atoms, we find that additional energy of the order of 3 kJ / g is required, and the total energy consumption should be $W_{las}$ ~ 20 kJ / g.

*6. 3. Irradiation parameters*

Consider the abilities of an infrared laser pulse at a distance $s_2 = 1000$ km from the irradiated body. For the diffraction divergence of the laser beam to be sufficiently small, it is necessary to integrate individual laser emitters into a laser array [6], similar to phase locking of individual emitters in a phased-array antenna.

Let the total diameter of the laser array be $d_{gr} = 3$m. Then the diffraction divergence angle will be $\theta_{dif} = \lambda / d_{gr} = 3 * 10^{-6}$, where $\lambda = 10$ μ is the wavelength of the laser radiation. Thus, at the distance $s_2 = 1000$ km the laser spot area can be estimated as $S = \pi * (s_2 * \theta_{dif})^2 = 30$ m$^2$.

Let the area of the quadrant on the rod to be evaporated for diverting the rod to the angle $\theta^\perp = p^\perp / (m_b * V_b) == 10^{-3}$ be $s_s = 3$ cm$^2$. In this case the geometric factor is $10^{-5}$. So, the laser energy required for heating and evaporating one gram of silicon is $W_{las1} = 20$ kJ. Considering the geometrical factor, the energy should be $10^5$ times higher, $W_{las2} = 2 * 10^9$ J, and radiation energy required to heat and evaporate 15 mg of silicon will be $W_{las3} = 30$ MJ.

Thus laser irradiation of a rod at a distance $s_2 = 1000$ km from the laser followed by evaporation of one of four quadrants on the rod will impart a transverse momentum $p^\perp = m_{jet} * V_{jet}$ to the rod, which will give rise to the deviation angle $\theta^\perp = p^\perp / (m_b * V_b) = =10^{-3}$. At a distance $s_1 = 100$ km from the nucleus of the comet this angle will result in diversion of the rod from the unperturbed trajectory by $\Delta l = s_1 * \theta^\perp = 100$ m.

**7. Rod penetration depth into the nucleus of a comet**

Let us find the rod penetration depth into the nucleus of the comet from the following considerations. First, we calculate the depth of penetration into aluminum, and then extend the results to the nucleus of the comet.

We take the density of aluminum to be $\rho_{Al} = 2.7$ g/cm$^3$ [5, p. 99]; the mass of 1 g-mol of aluminum is 27 g. The specific heat of aluminum is $c_{Al} = 24.35$ J / (mol * K) [5, p. 199], the melting point is $T_{mel} = 660$ C$^0$, the solid - liquid phase transition heat is $\Delta H_{mel} = 10.8$ kJ / mol [5, p. 289], and the boiling point is $T_{eva} = 2520$ $^0$C. To bring aluminum to a boil required 50 kJ / mol, the liquid - vapor phase transition is heat $\Delta H_{eva} = 293$ kJ / mol, [5, p. 289].

Adding all up, we find that evaporation of one mole of aluminum requires 375 kJ, and evaporation of one cubic centimeter of aluminum requires 37.5 kJ/cm$^3$. The cross sectional area of the rod is $S_{tr} = \pi d_{sh}^2 / 4 = 3.14 * 10^{-2}$ cm$^2$; thus, 1.2 kJ are needed for evaporation of an aluminum cylinder with a diameter equal to the diameter of the rod and a length of 1 cm.

The kinetic energy of the rod $mV_m^2 / 2$ at the velocity $V_m = 3$ km / s is $E_{kin} = 2.25 * 10^4$ J. From the data on reactions of meteorites with solids [7] it is known that at the velocity about 3 km / s, meteorites spend about 20% of their kinetic energy for evaporation of substance. In our case it means that is consumed by evaporation and 4.5 kJ will be spent for evaporation and the rod penetrates aluminum as deep as $l_{pen} = 4.5$ kJ / (1.2 kJ / cm) $\approx$ 3 cm.

The nucleus of the comet has a lower density than aluminum but on hitting it, the rod will certainly be destroyed, at least the transmitter will stop emitting electromagnetic energy.

It will thus be possible to determine whether the rod passed through the dust cloud surrounding the nucleus of the comet, or collided with the nucleus of the comet, in which case the rod will be destroyed and stop transmitting radio signals.

**8. Radio emission of rods**

Suppose that the transmitter attached to the rod emits radio waves at a wavelength $\lambda_{rad} = 3$ cm, (half -wave dipole) with a power $P_{rad} = 1$ μW. The parameters of the radiating pulse are taken to be the following: the pulse width $\tau_{pul} = 10$ μs and the repetition rate $F_{rep} = 100$ Hz. At a distance of R = 1000 km within the area of the receiving antenna $S_{ant}$ = 100 m$^2$ the power of the received signal will be $P_{res} = P_{rad} * S_{ant}/4\pi R^2 = 10^{-17}$ W.

The noise power of the receiver, if the receiver is a superconducting cavity with $T_{cav}$ = 1 K$^0$, will be $P_{noise} = kT_{cav}\Delta f$, where
k = 1.38 * 10$^{-23}$ J / degree is the Boltzmann constant, $T_{cav}$ = 1 K$^0$, $\Delta f_1 = 1 / \Delta \tau_{pul} = 10^5$ Hz is the reception band. Since every second there will come $F_{rep}$ = 100 Hz, the reception band can be taken to be two orders of magnitude narrower, $\Delta f_2 = 10^3$ Hz. This band corresponds to the resonator Q-factor $Q = f_{rad} / \Delta f_2 = 10^7$, where $f_{rad} = c / \lambda_{rad} = 10^{10}$ Hz is the frequency of the radiation. As a result, the received signal will exceed the energy noise by 3 orders of magnitude.

The Q-factor of modern superconducting resonators is of the order of $Q = 5 * 10^{10}$ [8]. This means that the intrinsic bandwidth of the resonator is a few hundredths of a hertz, $\Delta f_{cav} = f_0 / Q$, and it will have to be specially extended to allow appropriately fast processing of the received signals.

**9. Operation of the system equipment**

Figure 1 shows a diagram of the equipment.

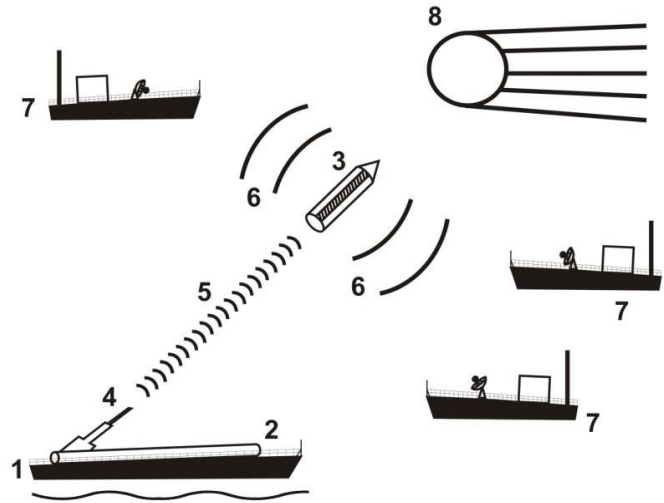

Fig.1. (1)- floating platform, (2)- accelerator, (3)- rods, (4)- four - channel IR laser, (5)- laser beam, (6)- radio waves emitted by a rod,
(7)- three independent stations receiving the emitted radio waves,
(8)- comet.

The equipment operates as follows.

The electromagnetic mass accelerator (2) are located on the floating platform (1). It accelerates rods (3) to the velocity $V_{in}$ = 6 km / s. Due to the pointed cone head and small asymmetry rods penetrate the Earth's atmosphere. The rods having an azimuthally and vertical speed approximately equal: $V_{\varphi, r} \approx$ 3.7 km / s. Rods carry four quadrants of silicon with different degrees of doping, so that each quadrant has its own resonant frequency of absorption. To control the flight path of a rod, it is irradiated (5) with four-channel infrared laser (4). Rods radiate radio waves (6) with the wavelength $\lambda_{rad}$ = 3 cm and power $P_{rad}$ = 1 µW. Radio waves are received by the receivers' with sensitivity $10^{-20}$ W, arranged on three floating platforms (7), determining the spatial coordinates of the rods. Rods pass through the dust cloud that hides the nucleus of the comet (8) without breaking. To meet with the nucleus of the comet, rods destroyed and cease to emit radio waves.

**Conclusions**

Receiving signals by three independent receivers can allow us to restore accurately the spatial coordinate of the emitting rods, and the presence or absence of the signal will allow us to judge whether the rod hit the nucleus of the comet or not. If rod, after passing through the dust cloud that hides the comet nucleus, continues to emit radio signals, it is precisely the area of the dust cloud. If after passing through the dust cloud the rod ceased to emit radio signals, it collided with the nucleus of the comet. Measuring the spatial coordinates of the rods that left the dust cloud hiding the nucleus of the comet and knowing the coordinates of the stopped rods, we can calculate the size of the comet nucleus.